\documentclass[prd, amsfonts, twocolumn, nofootinbib, showpacs]{revtex4-1}
\usepackage{amsmath}%
\usepackage{amsfonts}%
\usepackage{amssymb}%
\usepackage{graphicx}

\usepackage{bm}
\renewcommand{\vec}[1]{\bm{#1}}
\newcommand{\deriv}{\mathrm{d}}

\usepackage{color}

\begin{document}
\title{Real Space Approach to CMB deboosting}

\author{Amanda Yoho$^{1,2}$}
\author{Craig J. Copi$^{1}$}
\author{Glenn D. Starkman$^{1,2}$}
\author{Thiago S. Pereira$^{3,4}$}

\affiliation{$^{1}$
CERCA/ISO, Department of Physics, Case Western Reserve University,
10900 Euclid Avenue, Cleveland, OH 44106-7079, USA.}
\affiliation{$^{2}$ CERN, CH-1211 Geneva 23, Switzerland}
\affiliation{$^{3}$ Departamento de F\'isica, Universidade Estadual de Londrina
Campus Universit\'ario, 86051-990, Londrina, Paran\'a, Brazil}
\affiliation{$^{4}$ Institute of Theoretical Astrophysics, University of Oslo, 0315 Oslo, Norway}

\begin{abstract}
The effect of our Galaxy's motion through the Cosmic Microwave Background rest frame, which aberrates 
and Doppler shifts incoming photons measured by current CMB experiments, has been shown to produce
mode-mixing in the multipole space temperature coefficients. However, multipole space 
determinations are subject to many difficulties, and a real-space analysis can provide a straightforward 
alternative. In this work we describe a numerical method for removing 
Lorentz-boost effects from real-space temperature maps.
We show that to deboost a map so that one can accurately extract the temperature
power spectrum requires calculating the boost kernel at a finer pixelization than one might naively expect. In idealized cases that allow for easy comparison to analytic results, we have confirmed that there is indeed mode mixing among the spherical harmonic coefficients of the temperature. We find that using a boost kernel calculated at Nside=8192 leads to a 1\% bias in the binned boosted power 
spectrum at $\ell\sim 2000$, while individual $C_{\ell}$s exhibit $\sim$5\% fluctuations
around the binned average. However, this bias is
 dominated by pixelization effects and not the aberration and Doppler shift of CMB photons that causes
 the fluctuations. Performing analysis on maps with galactic cuts
does not induce any additional error in the boosted, binned power spectra over the full sky analysis. 
For multipoles that are free of resolution effects, there is no detectable deviation between the binned boosted and
unboosted spectra. This result arises because the power spectrum is a slowly 
varying function of $\ell$ and does {\it not} show that, in general, Lorentz boosts can be neglected for
other cosmological quantities such as polarization maps or higher-point functions.

\end{abstract}

\maketitle 

\section{Introduction}
Forthcoming results from the Planck satellite will set new  
limits on the primordial CMB temperature field at small angular scales. It will also strengthen 
our knowledge  about fluctuations at very large scales. In this regime, the temperature dipole 
is overwhelmingly the dominant contribution -- a fact usually attributed to the peculiar motion 
of the solar system, the Galaxy and the Local Group through the universe.
As pointed out in \cite{Kamionkowski:2002nd}, the very high amplitude of 
this dipole ($\sim100$ times larger than the primordial fluctuations of all higher multipoles) induces a 
peculiar quadrupole at a level detected by the WMAP satellite and accessible to the Planck satellite. 
When extracted from its primordial
counterpart, this information can be used to further constrain the physics of our motion with respect 
to the CMB.
We might also ask what are the contributions of this peculiar motion to the higher order 
multipole moments of the temperature power spectrum? This issue was first addressed in 
\cite{Challinor:2002zh}, where it was shown that \textit{if} the $C_\ell$s vary smoothly among neighboring 
$\ell$ modes in comparison to the aberration kernel that boosts the CMB sky, then the main effect is a small 
and unobservable modulation of the temperature power spectrum. Furthermore, it was also noted in 
\cite{Challinor:2002zh} that the multipolar convolution of the $a_{\ell m}$s with the aberration kernel 
has an expansion which is in powers of  $\beta\ell$, where $\beta\equiv v/c\sim10^{-3}$ is the Lorentz boost parameter. From 
this it follows that the expansion in harmonic space is not guaranteed to converge rapidly for 
$\ell\gtrsim800$, a region which most conspicuously concerns future measurements by Planck. Nevertheless, 
essentially all approaches devoted to deboost CMB maps were based on standard harmonic-space decompositions 
\cite{Challinor:2002zh,Kosowsky:2010jm,Chluba:2011zh,Amendola:2010ty}. 

More recently, it has been proposed \cite{Notari:2011sb} that one can deboost the CMB at the level of time 
ordered data. A follow-up paper showed that the power spectra derived 
from simulated aberrated maps affected cosmological 
parameter estimation~\cite{Catena:2012hq}. For their analysis they modified HEALPix code to 
produce a simulated, aberrated map from an input set of $C_{\ell}$s. This analysis did not 
include a way to translate between aberrated maps and non-aberrated maps.
Since CMB surveys directly measure time-ordered data which is converted to a temperature map, 
we will need a method for deboosting this temperature information 
for future analysis. In this work, we discuss our map deboosting method as
well as resolution limitations and give an estimate for appropriate map
resolutions for high-$\ell$ CMB analysis.

In this paper we assess the effect of a Lorentz boost on the temperature power spectrum through a full 
real-space analysis performed on temperature maps. There are  two relativistic effects to 
consider: a Doppler-shift of CMB photons and an aberration. The Doppler
shift of a map is determined the magnitude of the monopole and 
by the boost parameter. To  lowest order in $\beta$, the contribution is a  dipolar function multiplying the whole CMB 
map. At small angular scales the corrections are dominated by the aberration effect, 
which translates, in  real space, into a non-trivial distortion and overlap of CMB pixels. These effects 
in turn depend on both the resolution and the pixelization scheme used to cover the sphere. Given 
that Lorentz boosts do not preserve pixel shape, pixels with equal areas but different shapes are
distorted differently. Moreover, the overlap of boosted pixels with unboosted ones
depends heavily  on the tessellation of the sphere. Using the spherical
projection pixelization scheme described in \cite{Gorski:2004by}, which 
maintains straight line boundaries between equatorial and polar cap pixels and is easily related
to the traditional HEALPix equal area pixelization \footnote{http://healpix.jpl.nasa.gov}, we 
construct a real-space matrix that fully accounts for  
the pixel distortions and overlaps among pixels. Our approach not only circumvents complications associated with 
an expansion in $\beta$ altogether, but also benefits from the already developed real-space tools used in 
standard CMB analysis.

We organize the paper as follows: in Section \ref{boost_effect} we review the basics of
boosting the CMB in harmonic space and stress its limitations. In Section \ref{real_space} we 
present our general real-space approach and discuss its consequences when applied to the HEALPix 
pixelization scheme. We apply our method to simulated full sky maps and compare the resulting power 
spectrum in Section 
\ref{real_space}, we will describe the effect of map resolution on deboosting and present limitations 
in~\ref{res}, and we present concluding remarks in section \ref{discussion}.

\section{Lorentz boost effects on CMB photons}\label{boost_effect}
Relativistic effects have been outlined in \cite{mckinley1980relativistic}, here we summarize
relevant points and and describe our notation.

If an observer in the rest frame of the CMB (denoted $S$) measures a  photon of frequency
$\nu$ arriving along a line of sight $\hat{\vec{n}}$, then an observer in another frame,
$S^{\prime}$, that is moving with respect to the CMB at velocity $v\hat{\vec{v}}$
will measure the incoming photon to be arriving along a different line-of-sight,
$\hat{\vec{n}}^{\prime}$, with a different frequency, $\nu^\prime$. (Note that we will not
concern ourselves here with any ambiguities in determining $S$ associated with the existence
of inhomogeneities,  in particular a cosmological dipole, on the assumption that that
dipole is $\mathcal{O}(10^{-5})$ and smaller than the effects we will uncover.) The motion of
the observer in $S^{\prime}$ thus induces two effects: a Doppler shift  in the photon
frequency and an aberration  -- a shift in the direction from which the photon arrives.
These two effects can be seen explicitly in the relation between $\hat{\vec{n}}$ and
$\hat{\vec{n}}^{\prime}$
\begin{equation}
\label{aberration}
\hat{\vec{n}}^{\prime}=\left(\frac{\cos\theta+\beta}{1+\beta\cos\theta}\right)\hat{\vec{v}}
+\frac{\hat{\vec{n}}-\hat{\vec{v}}\cos\theta}{\gamma (1+\beta \cos\theta )},
\end{equation}
where $\beta\equiv v/c$ and $\cos\theta\equiv\hat{\vec{n}}\cdot \hat{\vec{v}}$.
The change in observed frequency in $S^{\prime}$ is given by a simple Lorentz
transformation 
\begin{equation}\label{doppler}
\nu^{\prime}=\gamma \nu (1+\beta \cos\theta) \,,
\end{equation}
where $\gamma=(1-\beta^{2})^{-1/2}$ is the standard Lorentz factor. This angle $\theta$ is related to the angle
$\theta^{\prime}$, measured in the frame 
$S^{\prime}$, via 
\begin{equation}\label{angle}
\cos\theta' = { \frac{\cos\theta + \beta}{1+ \beta\cos\theta}} \, .
\end{equation}

For small $\beta$ one can clearly expand $\cos\theta'$ in a convergent series in $\beta$ around $\cos\theta$.
It would appear obvious that one can therefore do the same for the spherical harmonic function $Y_{\ell m}(\theta',\phi) \propto P_\ell^m(\cos\theta')$.
However, this is not the case, the correct expansion parameter for $Y_{\ell m}$ is $\beta\ell$.  
This expansion will therefore break down (or at least converge poorly) when $\ell\gtrsim1/\beta$. 
(This becomes obvious when one recalls that $P_\ell(\cos\theta)$ has ${\cal O}(\ell)$ zeros between $-1$ and $+1$,
thus changing $\cos\theta$ by ${\cal O}(\beta)$ moves one zero past the next for $\ell\gtrsim2/\beta$. )
This issue was originally noticed in \cite{Challinor:2002zh} where it was 
indeed recognized that this convergence breakdown leads to severe difficulties in evaluating the transformation of 
a spherical harmonic expansion of the sky at  $\ell\gtrsim1/\beta$. 
Nonetheless, they claimed that if the $C_\ell$s are smooth functions of $\ell$ compared to  the 
transformation kernel,  then this poor convergence is a red herring, and  to second order in $\beta$:
\begin{equation}
C'_\ell\approx C_\ell(1+4\beta^2+\mathcal{O}(\beta^3))\,.
\end{equation}

We want to stress here that these assumptions are idealized, but far from
ideal.  In practice we do not measure the theoretical power spectrum, only
its estimator $\langle\mathcal{C}_\ell\rangle$. Real data is contaminated
from many different sources which destroys the assumption of
smoothness. This suggests a real-space approach is more appropriate since
the aberration can be accounted for in terms of pixel distortions without
the need to resort to a $\beta\ell$ expansion. We show in the next
section how this can be implemented.

\section{The Real Space Approach to Boost Corrections}
\label{real_space}
We wish to transform the boosted, pixelized temperature map into an unboosted map that can be
analyzed with techniques already available for CMB analysis. To do this we  need to compute a 
boost matrix that characterizes how the unboosted pixels transform into boosted ones. 

The pixelized
representation of the temperature fluctuation $\Delta T(\hat{\vec{n}})$ can be defined by:
\begin{equation}
\label{average}
\Delta T(p)\equiv\frac{1}{\Omega_{p}}\int_{p} \Delta T(\hat{\vec{n}}) \,\deriv^{2}\hat{\vec{n}},
\end{equation}
where the integral is over the pixel $p$ which has area $\Omega_{p}$.
According to this definition, the value at pixel $p'$ of the boosted
fluctuation $\Delta T'(\hat{\vec{n}}')$ will be: 
\begin{eqnarray}
\Delta T'(p') & = & \frac{1}{\Omega_{p'}}\int_{p'}\Delta T'(\hat{\vec{n}}') \,\deriv^{2}\hat{\vec{n}}'
\label{pix_old}\\
 & = & \frac{1}{\Omega_{p'}}\int_{p'} D^{-1}(\hat{\vec{n}})\Delta
T(\hat{\vec{n}}) \, \deriv^{2}\hat{\vec{n}}
\label{pix_new}
\end{eqnarray}
where we have made use of the following transformations~\cite{mckinley1980relativistic}:
\begin{eqnarray}
\Delta T'(\hat{\vec{n}}') & = & D(\hat{\vec{n}})\Delta T(\hat{\vec{n}}) \\
\deriv^{2}\hat{\vec{n}}' & = & D^{-2}(\hat{\vec{n}})\,\deriv^{2}\hat{\vec{n}}\\
D(\hat{\vec{n}}) & \equiv & \gamma(1+\beta\hat{\vec{v}}\cdot\hat{\vec{n}}).
\end{eqnarray}
Note that when going from (\ref{pix_old}) to (\ref{pix_new}), the boundary 
of integration has not changed. In fact, the integral 
over $p'$ should be replaced by a sum of integrals over all pixels overlapping with $p'$. Note also
that the overlaps can (and do) occur among pixels that are not neighbors of the original pixel. We can therefore write
\begin{equation}
\Delta T'(p') = \sum_p \Lambda_{pp'}\, D^{-1}(p)\Delta T(p),
\label{transf}
\end{equation}
where $\Lambda_{pp'}$ are matrix coefficients encoding the fraction overlapping of pixels. This transformation includes the pixelized versions of both
 Doppler, $D^{-1}(p)$,
and aberration, $\Lambda_{pp'}$, effects, and can be implemented entirely in pixel-space. The 
Doppler term is an overall dipole and significantly contributes only at
very large scales. The boost matrix coefficients, 
$\Lambda_{pp'}$, on the other hand, are a function of pixels $p$ and $p'$ and strongly depend on the
aberration effect. They contain the information 
about the fractional overlap of an unboosted pixel $p$ with a boosted pixel $p^{\prime}$.

To compute the boost matrix, we make use of the spherical
projection of the HEALPix sphere described in \cite{Gorski:2004by}. In this
pixelization the pixel boundaries are straight lines in both the equatorial
and polar-cap regions which makes the pixels easier to boost.
To compute the matrix, we find which 
new pixel the original pixel center is boosted to and search its nearest neighbors and compute
the overlap area of the boosted pixel using an approximation of the exact area integral. These 
fractional area overlaps comprise the boost matrix.

\section{Comparison with exact results}
\label{al0}

To get a qualitative understanding of how a power spectrum is transformed under a Lorentz boost,
we simulated skies with power in only a single mode with $\ell,\, m=200,\, 0$.
We boosted this sky with $\beta =0.01$, so that significant mode mixing would set in around $\ell\sim 100$, 
 and extracted the resulting $a_{\ell^{\prime}m^{\prime}}$. The 
 analytic calculation of the integral for the boosted $a^{\prime}_{\ell m}$~\cite{Pereira:2010dn} is tractable for
 this choice of multipole, and provides exact results for comparison to our numerical results.

\begin{figure}
\includegraphics[scale=0.35]{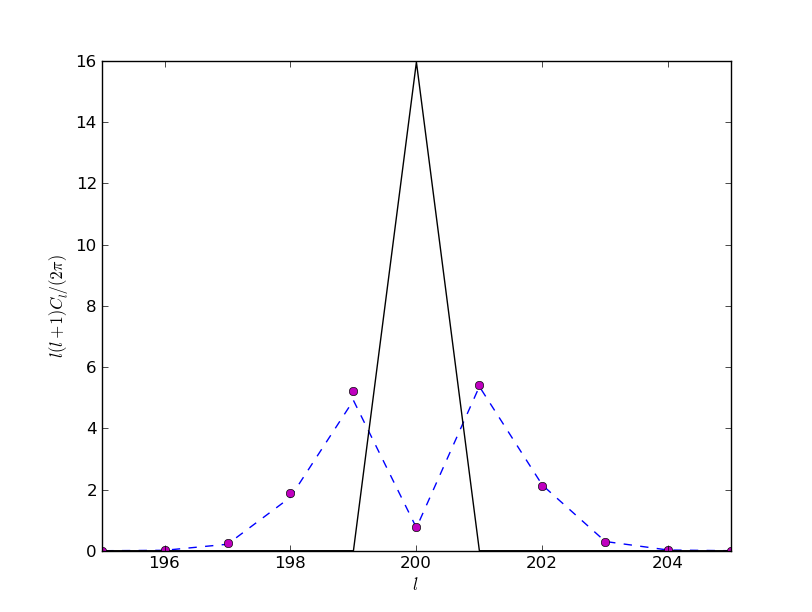}
\caption{Comparison of boosted (blue dashed line) and unboosted (black solid line) temperature power spectra from a simulated
sky at Nside=512 with all power in the $\ell =200$, $m=0$ mode to the exact solution. 
Differences between the numerically boosted result and the exact solution (magenta
dots) are predominantly due to map resolution effects. 
A $\beta$ value of .01 was used to obtain these results.}
\label{al0_fig}
\end{figure}

Figure~\ref{al0_fig} clearly shows  
that power is shifted from the mode which contained all of the power in the unboosted sky 
to nearby multipoles in the boosted sky.
Additionally, we see that the analytic results are in good agreement with the numerical values
for an Nside=512 map.
The discrepancies between the numerical and analytic calculations can be attributed to 
errors induced by pixelization of the temperature map. We will discuss this
effect at length in the following section.

\section{Resolution Effects}
\label{res}

We have found that the resolution of the input map used for deboosting has a direct effect on the 
resultant power spectrum. This is due to pixelization effects -- the inaccuracies inherent in replacing
the true mapping between points on the sphere that the boost represents, with mappings from pixels
of the unboosted sky, to pixels of the boosted sky.
The fractional overlaps contributing to a pixel  average inevitably includes area that the original, boosted map should not include. 
It is therefore important to check that deboosting is  done at a suitable resolution for whichever range of $\ell$ is to be included 
in the subsequent analysis.   Higher  multipoles are most susceptible to pixelization-induced effects. 
If the map resolution is sufficient, then after a map has been deboosted, 
an equal boost in the  opposite direction should reproduce the original power spectrum over the
full range of $\ell$ to the desired/required accuracy.
 This can be used to check for  residual resolution effects in the map at any preferred scale. 

\begin{figure}
\includegraphics[scale=0.35]{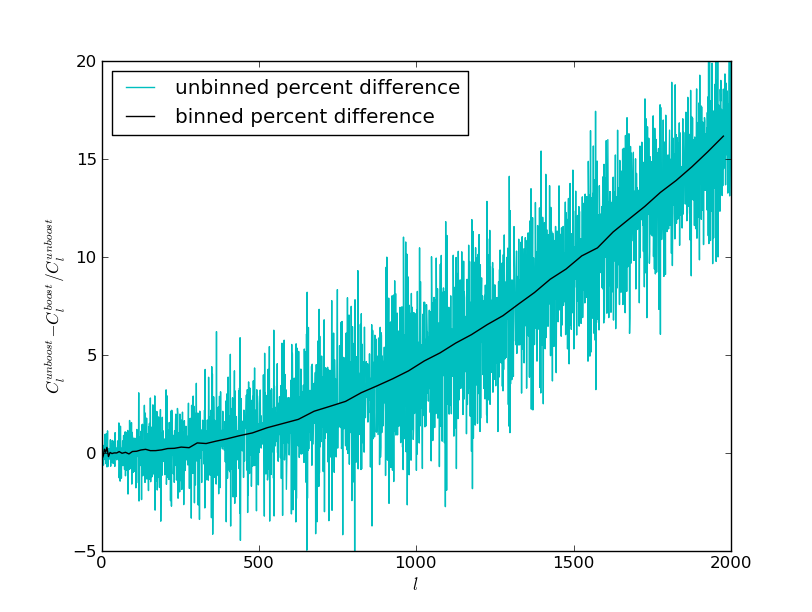}
\includegraphics[scale=0.35]{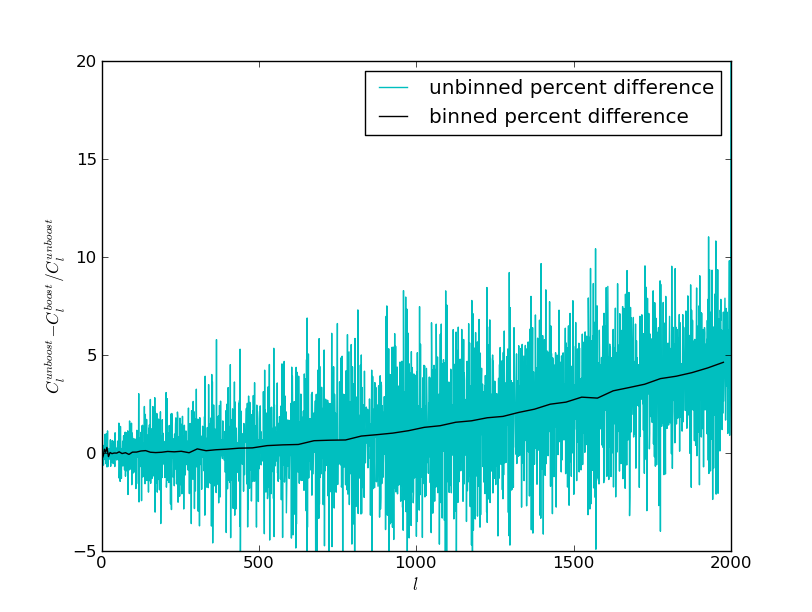}
\includegraphics[scale=0.35]{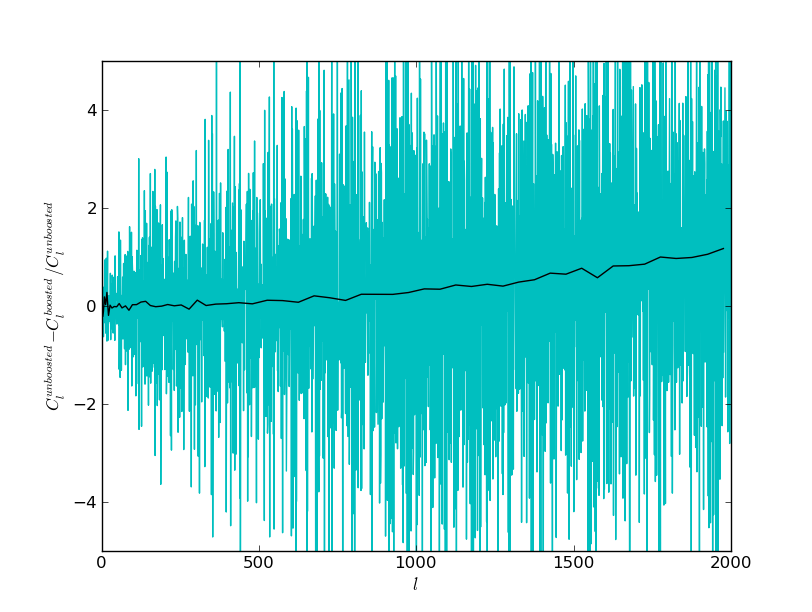}
\caption{Comparison of power spectra produced from boosted versus unboosted full-sky maps at 
Nside=2048(top) 4096(middle) and 8192(bottom). The blue line shows the difference for 
individual $C_{\ell}$s and the black line shows the difference in binned $C_{\ell}$s. Values are quoted in percent.}
\label{fullsky}
\end{figure}

\begin{figure}
\includegraphics[scale=0.35]{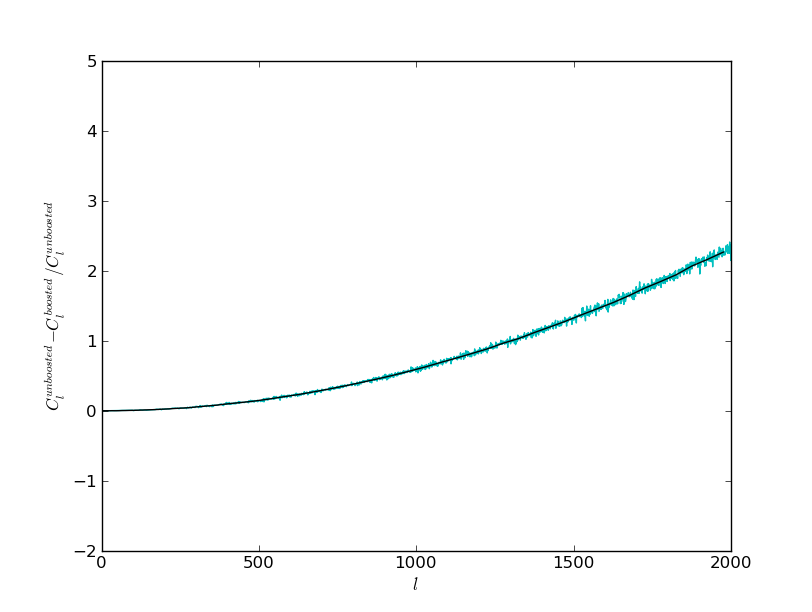}
\caption{Comparison of power spectra produced from a map that was not boosted versus a 
map that had been boosted and deboosted. The difference should be zero, as deboosting should
return the map pixels to their original values. The difference shown is therefore due to pixelization 
effects. The blue line shows the difference for 
individual $C_{\ell}$s and the black line shows the difference in binned $C_{\ell}$s. Values are quoted in percent.}
\label{deboost}
\end{figure}

Figure \ref{fullsky} shows how resolution effects 
change the power spectrum for maps at resolutions of Nside=2048, 4096, and 8192, which are higher
than those traditionally used for CMB analysis. We show that in order to circumvent 
the problem one must overpixelate the boost kernel for accurate deboosting.
The Nside=8192 map, which has 
been deboosted with a boost kernel calculated at Nside=8192, has a 1\% difference between the unboosted and boosted power spectrum for $\ell\sim 2000$.
We confirm that this is a resolution based effect, since an Nside=8192 map which has 
been boosted then deboosted has a 2\% residual difference in the  power spectrum for $\ell\sim 2000$
as seen in Figure \ref{deboost},
in agreement with the conclusion that the 1\% inaccuracy in the power spectrum in Fig.~\ref{fullsky}
is predominantly due to resolution effects. Interestingly, the fluctuations in the individual 
$C_{\ell}$s are small for the deboosted map, suggesting that the $\sim 5\%$ fluctuations of 
boosted $C_{\ell}$s around the binned bias in Fig.~\ref{fullsky} are {\it not} due to pixelization
effects.

These resolution effects can clearly lead to inaccuracies in the  extracted cosmological
parameters if the difference in produced power spectra are assumed to be primordial in origin
rather than due to the data processing. 

\begin{figure}
\includegraphics[scale=0.35]{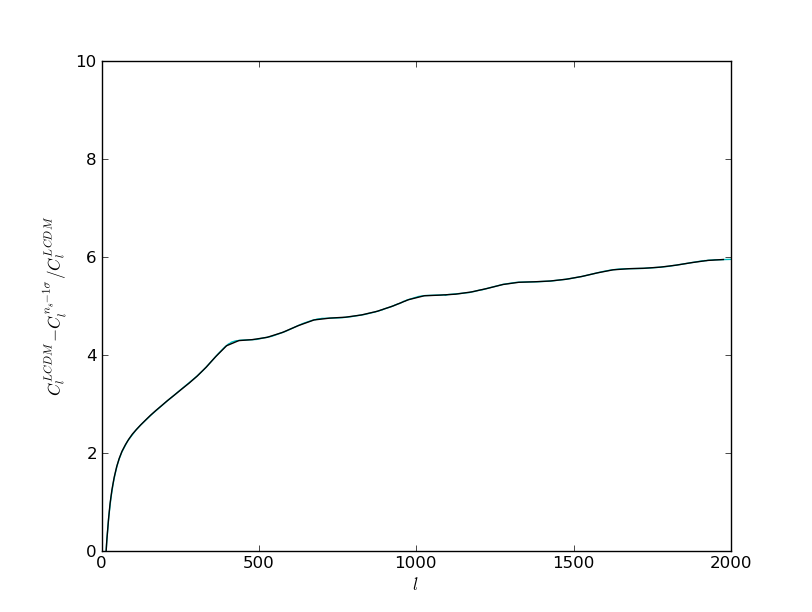}
\caption{Percent difference between a power spectrum generated by CAMB~\cite{Lewis:1999bs} with 
$n_{s}-1\sigma$ and a standard LCDM power spectrum. Values are quoted in percent.}
\label{ns}
\end{figure}

We show in Figure \ref{ns} the percent difference between two 
power spectra that have different
values of $n_{s}$ (one was produced with the WMAP 7-year best-fit value, the other power
spectrum was produced with the best-fit value plus one sigma using CAMB \footnote{http://camb.info},\cite{Lewis:1999bs} ), showing that a difference in the 
spectrum at large scales gives a similar difference as the resolution effects at Nside=4096. This suggests
that resolution effects could be confused with changes in cosmological parameters unless one is
careful to deboost with the appropriate boost kernel. As shown in Fig. \ref{fullsky}, an Nside=8192 boost
kernel would be sufficient to see changes in parameters that result from deviations larger
than 1\% in the binned power spectrum.

A recent paper by Catena and Notari~\cite{Catena:2012hq} claimed that $n_{s}$ extracted from 
simulated maps at a resolution of Nside=2048 with intrinsic aberration and WMAP-like noise differed by
approximately 1 sigma from value extracted from an unaberrated map.

\begin{figure}
\includegraphics[scale=0.35]{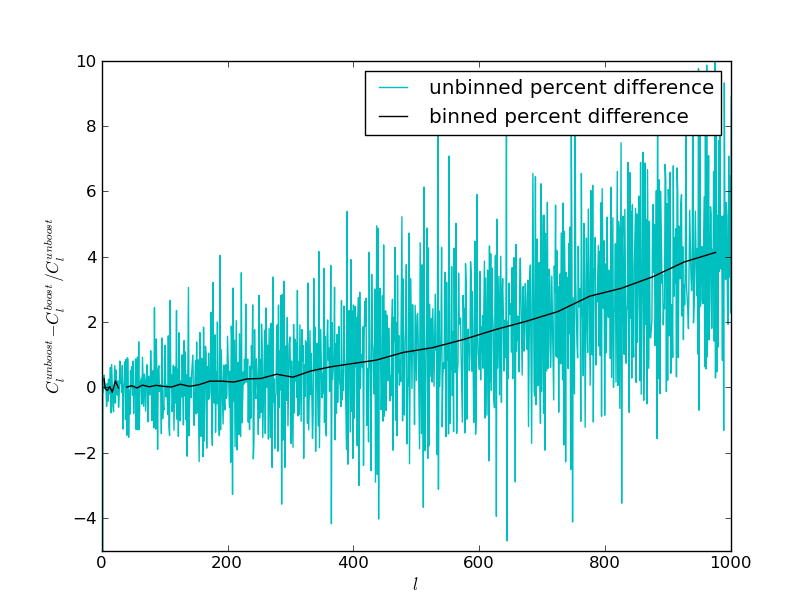}
\includegraphics[scale=0.35]{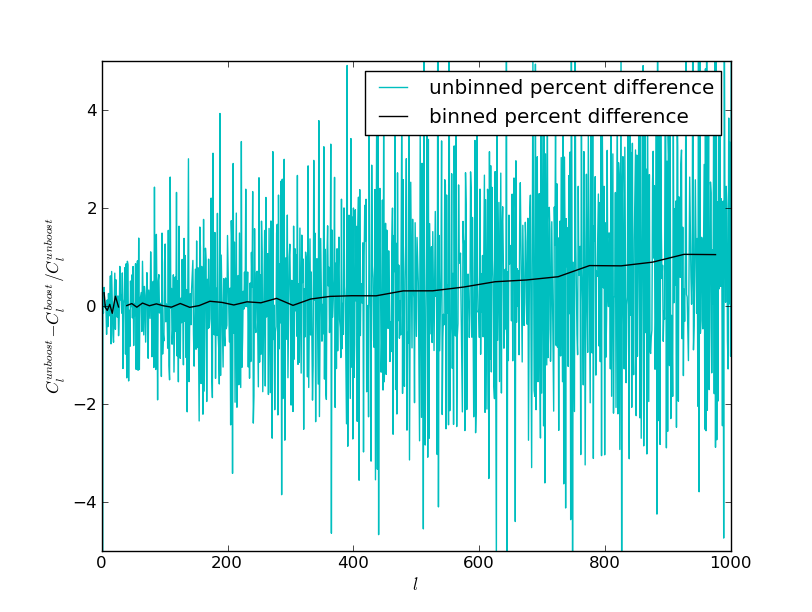}
\caption{Comparison of Nside=2048(top) and Nside=4096(bottom). These have been boosted and downgraded to Nside=512. Values are quoted in percent.}
\label{dg1}
\end{figure}

\begin{figure}
\includegraphics[scale=0.35]{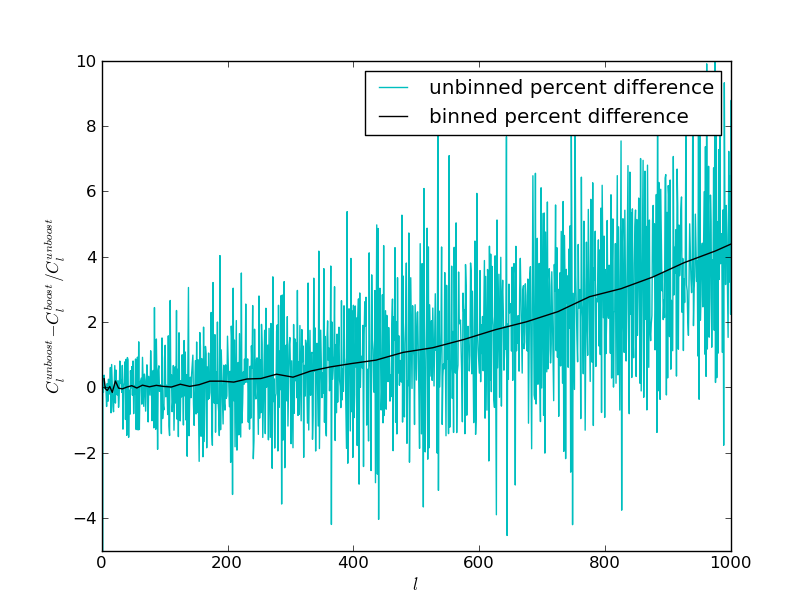}
\includegraphics[scale=0.35]{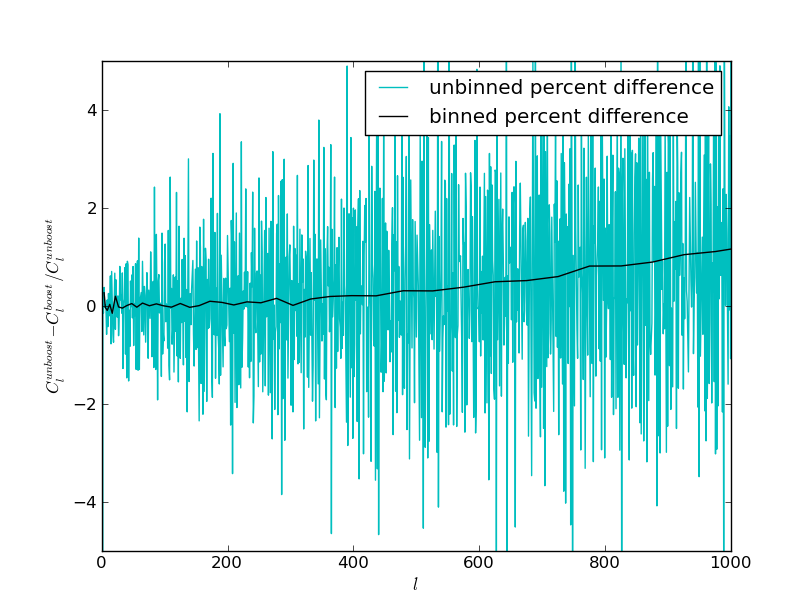}
\caption{Comparison of Nside=2048(top) and Nside=4096(bottom). These have been boosted and downgraded to Nside=1024. Values are quoted in percent.}
\label{dg2}
\end{figure}

We also note that boosting at a high resolution and downgrading the map before extracting the 
power spectrum gives better results than simply boosting at the desired resolution from the onset
as seen in Figures \ref{dg1} \& \ref{dg2}.
Therefore one should calculate their boost kernel at a pixelization well above the map pixelization before applying it to a map.

\section{Analysis using masked maps}
\label{mask}

\begin{figure}
\includegraphics[scale=0.35]{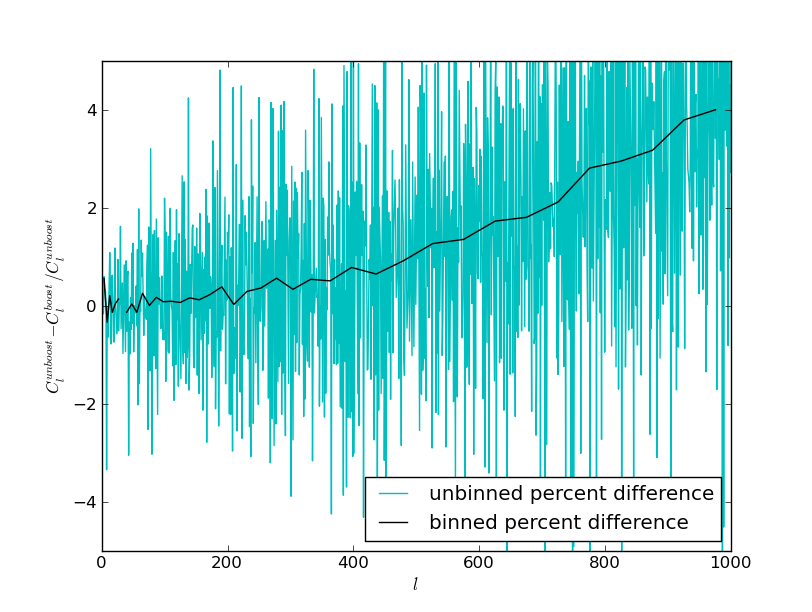}
\includegraphics[scale=0.35]{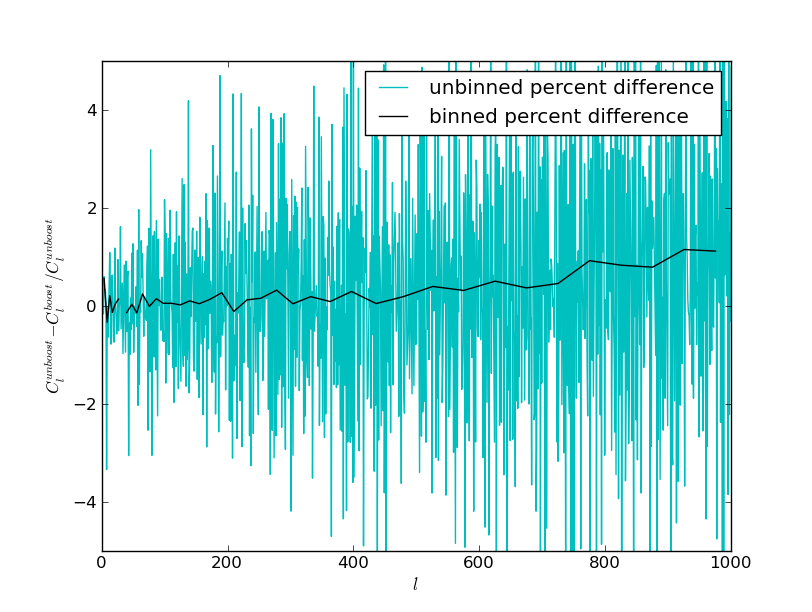}
\includegraphics[scale=0.35]{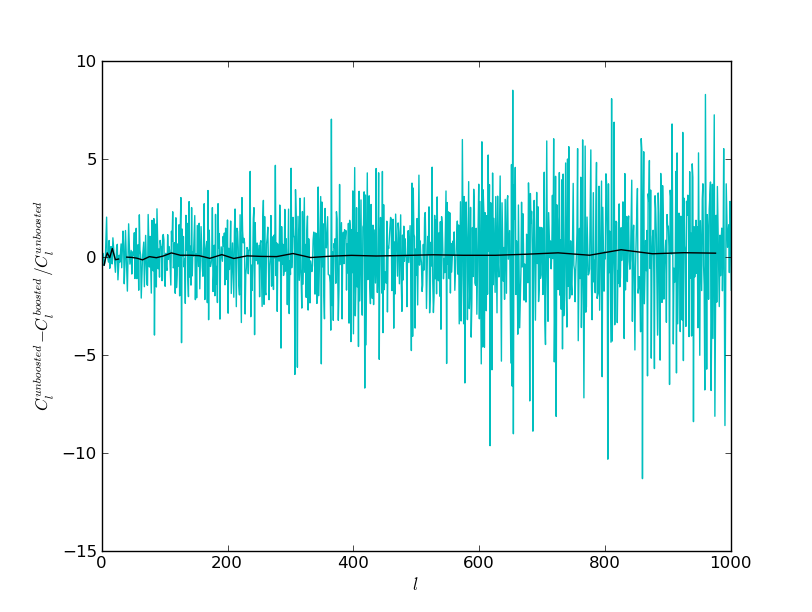}
\caption{Comparison of nside=2048(top), 4096(middle), and 8192(bottom). These have been boosted, downgraded to nside=512, and masked with a galactic mask provided by LAMBDA. Values are quoted in percent.}
\label{galcut1}
\end{figure}

\begin{figure}
\includegraphics[scale=0.35]{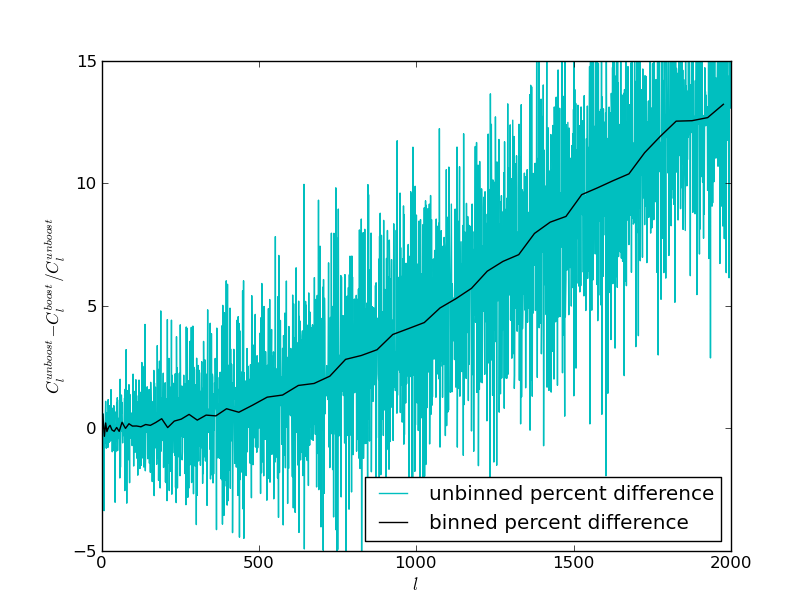}
\includegraphics[scale=0.35]{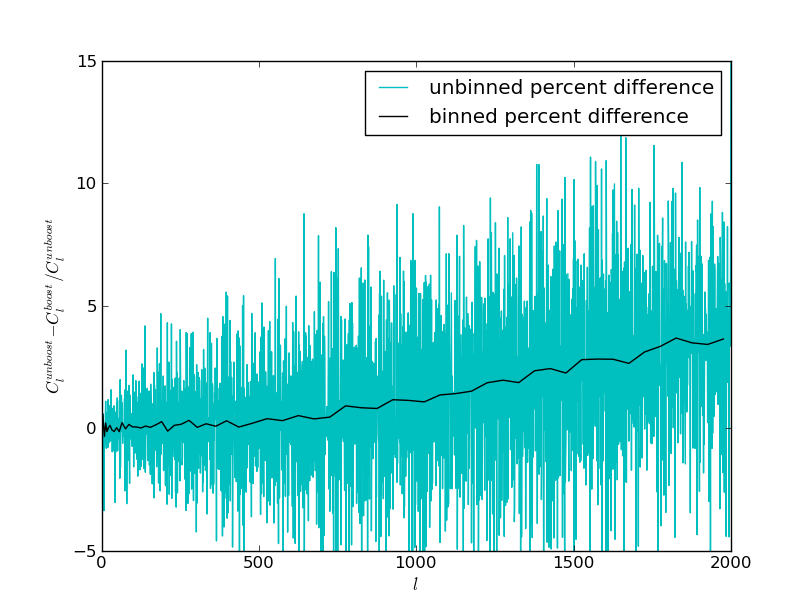}
\includegraphics[scale=0.35]{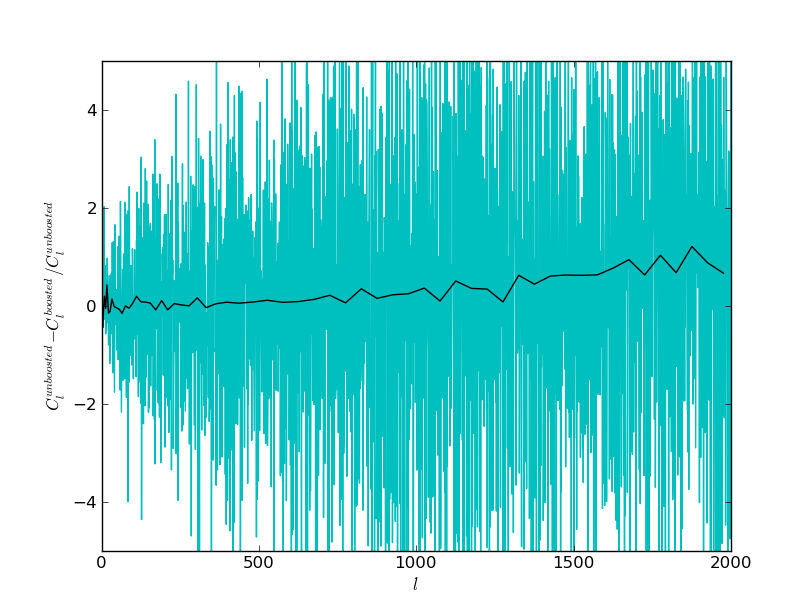}
\caption{Comparison of nside=2048(top), 4096(middle), and 8192(bottom). These have been boosted, downgraded to nside=1024, and masked with a galactic mask provided by LAMBDA. Values are quoted in percent.}
\label{galcut2}
\end{figure}

Because masking a map is known to correlate modes at all scales~\cite{Hinshaw:2003ex}, we wanted
to investigate the effect of both boosting and masking a temperature map on the extracted
power spectrum. To that end, we employed two methods. We used boosted maps at Nside
of 2048, 4096, and 8192, rotated them to the direction of the dipole, downgraded them to 
both Nside=512 and 1024 (resolutions traditionally used for CMB analysis), and masked then using 
the conservative galactic mask released by WMAP
from the LAMBDA website~\footnote{http://lambda.gsfc.nasa.gov}. The same was done for unboosted maps for a comparison
between the resultant power spectra shown in Figure \ref{galcut1} for Nside=512 and 
Figure \ref{galcut2} for Nside=1024. We confirmed before doing full mask analysis that rotating
each map did not contribute to our results.

\begin{figure}
\includegraphics[scale=0.35]{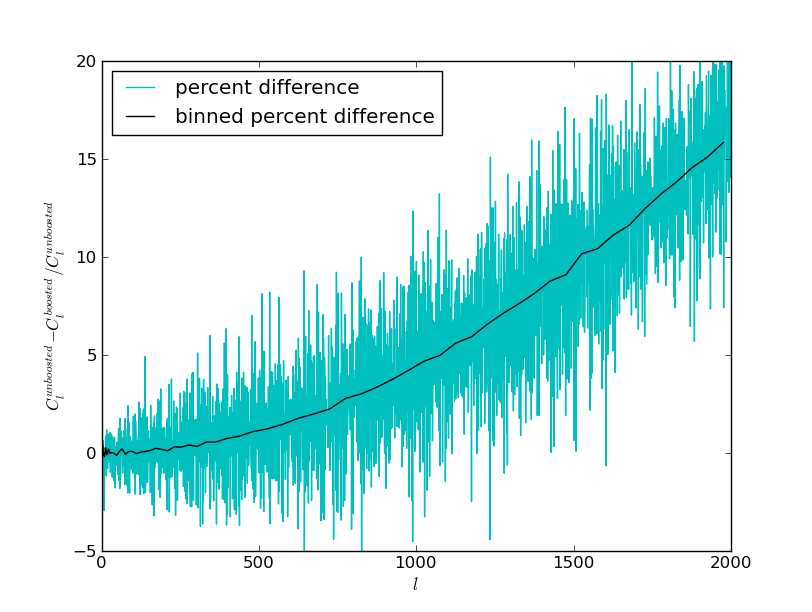}
\includegraphics[scale=0.35]{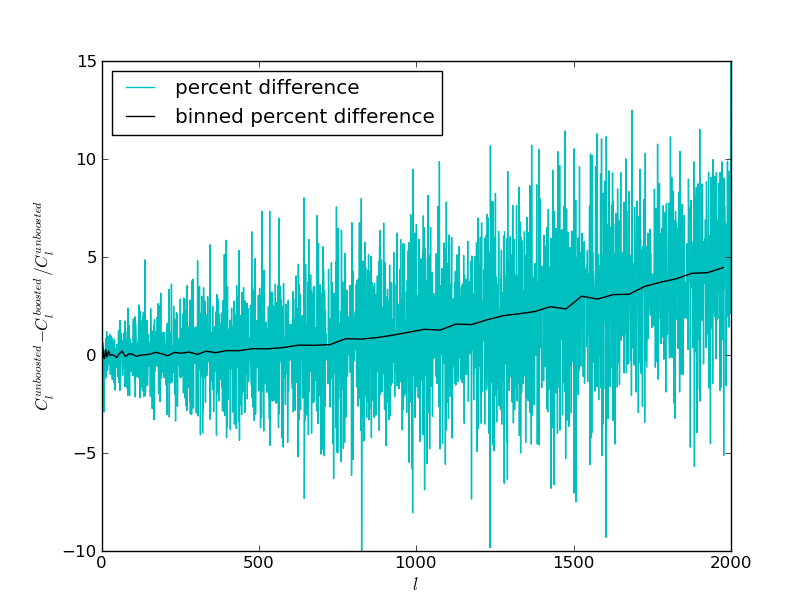}
\caption{Comparison of nside=2048(top) and nside=4096(bottom). These have been boosted and masked with
a 20 degree isolatitude equatorial strip. Values are quoted in percent.}
\label{iso}
\end{figure}

We also directly masked Nside 2048 and 4096 maps with a 20 degree wide isolatitude galactic cut.
This was done because the released galactic masks are not produced at resolutions higher than 1024. The
results are shown in Figure \ref{iso}.

For the high resolution maps with isolatitude masks, 
we have found that masking a map produces
a negligible effect on the power spectrum extracted from a boost map. The most noticeable difference
is that individual $C_{\ell}$s oscillate by a wider range than the unmasked power spectra, but 
the binned differences between masked and unmasked are nearly identical.

For the maps that have been masked with a galactic cut, we see that boosting at a much higher Nside, 
downgrading the map and then masking at either Nside=512 or 1024 give more accurate results than
boosting and masking at the same resolution.

\section{Discussion}\label{discussion}

We have investigated a real-space approach to deboosting a CMB temperature map by calculating
fractional area overlaps of boosted pixels with the HEALPix grid in the spherical projection. We have
shown that the resolution at which the deboosting is performed greatly affects the resulting power spectrum
and therefore one should take great care when extracting quantitative information from deboosted maps.
We have shown that for a boost kernel that is calculated at Nside=8192, the highest resolution used in this work, there is a 1\% difference in the binned angular power spectrum at $\ell =2000$
between full-sky boosted and unboosted maps. Because boosting, rotating the map 180 degrees, and boosting
again does not return the power spectrum to its unboosted value but instead further decreases the power
at large multipoles at a 2\% level, the behavior is dominated by resolution effects and is not predominantly
physical in origin. 

We have shown that resolution effects could lead one to incorrectly 
conclude that cosmological quantities were biased by this systematic, as power spectra generated
from CAMB with 1 sigma difference in $n_{s}$ and the WMAP best fit value of $n_{s}$ have percent
differences at large multipoles consistent with those we see from boosted and unboosted maps
at Nside=4096.

Since masking maps mixes multipoles, we compared using masked maps that had boost
contributions included to using unboosted masked maps for power spectrum calculations. We found 
that, while masking along with boosting caused the difference between each individual $C_{\ell}$ to vary
more than for full-sky maps, masking boosted maps had little effect on the differences between binned
$C_{\ell}$ values. We found that if one wants to deboost a map and later mask with the
publicly available galactic cuts released by WMAP (which can be downloaded at Nside=512 or Nside=1024)
one should first deboost with a higher resolution boost kernel before masking the map.

But, as Section~\ref{al0} shows, Lorentz boosts do move power to nearby multipoles -- the 
behavior we see here is a feature of our {\it particular} power spectrum and the fact that it is
a slowly varying function. It is not a generic statement about Lorentz boosts for any cosmological quantity
and highlights a need for polarization fields and the three-point function to be investigated.

\begin{acknowledgments}
We would like to thank Anthony Challinor, Arthur Kosowsky, Raul Abramo, Wayne Hu, Kendrick Smith, 
Maik Stuke, Dominik J. Schwarz, and Pascal Vaudrevange for useful conversations 
during the preparation of this work. TSP thanks the physics department of Case Western Reserve University 
and the Institute of Theoretical Astrophysics at Oslo for their hospitalities during the stages of this work. 

GDS and AY are supported by a grant from the US Department of Energy and were supported early in this
 work by NASA under cooperative agreement NNX07AG89G. AY is supported by NASA Headquarters under the 
 NASA Earth and Space Science Fellowship Program - Grant TRN507323.

We acknowledge the use of the Legacy Archive for Microwave Background Data Analysis (LAMBDA). Support for LAMBDA is provided by the NASA Office of Space Science.

\end{acknowledgments}



\begin{thebibliography}{99}  

 
\bibitem{Kamionkowski:2002nd} 
  M.~Kamionkowski and L.~Knox,
  ``Aspects of the cosmic microwave background dipole,''
  Phys.\ Rev.\ D {\bf 67}, 063001 (2003)
  [astro-ph/0210165].

  
\bibitem{Challinor:2002zh} 
  A.~Challinor and F.~van Leeuwen,
  ``Peculiar velocity effects in high resolution microwave background experiments,''
  Phys.\ Rev.\ D {\bf 65}, 103001 (2002)
  [astro-ph/0112457].
  
\bibitem{Kosowsky:2010jm}
  A.~Kosowsky, T.~Kahniashvili,
  ``The Signature of Proper Motion in the Microwave Sky,''
  [arXiv:1007.4539 [astro-ph.CO]].
  
\bibitem{Amendola:2010ty}
  L.~Amendola, R.~Catena, I.~Masina {\it et al.},
  ``Measuring our peculiar velocity on the CMB with high-multipole off-diagonal correlations,''
  [arXiv:1008.1183 [astro-ph.CO]].

  
\bibitem{Chluba:2011zh} 
  J.~Chluba,
  ``Aberrating the CMB sky: fast and accurate computation of the aberration kernel,''
  arXiv:1102.3415 [astro-ph.CO].
  
\bibitem{Notari:2011sb} 
  A.~Notari and M.~Quartin,
  JCAP {\bf 1202}, 026 (2012)
  [arXiv:1112.1400 [astro-ph.CO]].
  
  
\bibitem{Catena:2012hq} 
  R.~Catena and A.~Notari,
  ``Cosmological parameter estimation: impact of CMB aberration,''
  arXiv:1210.2731 [astro-ph.CO].

\bibitem{Gorski:2004by} 
  K.~M.~Gorski, E.~Hivon, A.~J.~Banday, B.~D.~Wandelt, F.~K.~Hansen, M.~Reinecke and M.~Bartelman,
  Astrophys.\ J.\  {\bf 622}, 759 (2005)
  [astro-ph/0409513].
  
    
\bibitem{mckinley1980relativistic}
  McKinley, J.M.,
  ``Relativistic transformation of solid angle'',
  American Journal of Physics, {\bf 48}, 612--614, 1980.
  
  
\bibitem{Pereira:2010dn}
  T.~S.~Pereira, A.~Yoho, M.~Stuke {\it et al.},
  ``Effects of a Cut, Lorentz-Boosted sky on the Angular Power Spectrum,''
  [arXiv:1009.4937 [astro-ph.CO]].

\bibitem{Lewis:1999bs}
     A. Lewis, A. Challinor, A. Lasenby, Anthony,
    ``Efficient Computation of {CMB} anisotropies in closed {FRW} models'',
     Astrophys. J., {\bf 538}, 473-476, 2000.
     astro-ph/9911177".


\bibitem{Hinshaw:2003ex} 
  G.~Hinshaw {\it et al.}  [WMAP Collaboration],
  ``First year Wilkinson Microwave Anisotropy Probe (WMAP) observations: The Angular power spectrum,''
  Astrophys.\ J.\ Suppl.\  {\bf 148}, 135 (2003)
  [astro-ph/0302217].
 
  
\end{thebibliography}
\end{document}